\documentclass[twocolumn]{aa}
\usepackage{graphicx}
\usepackage{txfonts}
\begin{document}
   \title{Effective temperatures and radii of planet-hosting stars from 
          IR photometry}

   \author{I. Ribas\inst{1}
          \and
          E. Solano\inst{2}
          \and
          E. Masana\inst{1}
          \and
          A. Gim\'enez\inst{3}}

   \offprints{I. Ribas}

   \institute{Departament d'Astronomia i Meteorologia, Av. Diagonal, 647,
              08028 Barcelona, Spain\\
              \email{iribas@am.ub.es, emasana@am.ub.es}
         \and
              Laboratorio de Astrof\'{\i}sica Espacial y F\'{\i}sica
              Fundamental (LAEFF), Apdo. 50727, 28080 Madrid, Spain \\
              \email{esm@laeff.esa.es}
         \and
              Research and Scientific Support Department, ESA, ESTEC, Postbus
              299, 2200 AG Noordwijk, The Netherlands\\
             \email{agimenez@rssd.esa.int}
             }

   \date{Received; accepted}

   \abstract{In this paper we present and analyse determinations of
effective temperatures of planet-hosting stars using infrared (IR)
photometry.  One of our goals is the comparison with spectroscopic
temperatures to evaluate the presence of systematic effects that could
alter the determination of metal abundances. To estimate the stellar
temperatures we have followed a new approach based on fitting the observed
2MASS IR photometry with accurately calibrated synthetic photometry.
Special care has been put in evaluating all sources of possible errors and
incorporating them in the analysis. A comparison of our temperature
determinations with spectroscopic temperatures published by different
groups reveals the presence of no systematic trends and a scatter
compatible with the quoted uncertainties of 0.5--1.3\%. This mutual
agreement strengthens the results of both the spectroscopic and IR
photometry analyses. Comparisons with other photometric temperature
calibrations, generally with poorer performances, are also presented. In
addition, the method employed of fitting IR photometry naturally yields
determinations of the stellar semi-angular diameters, which, when combined
with the distances, results in estimations of the stellar radii with
remarkable accuracies of $\sim$2--4\%. A comparison with the only star in
the sample with an empirically determined radius (HD 209458 -- from
transit photometry) indicates excellent agreement.
   }

   \maketitle
%

\section{Introduction}

The characterization of the properties of planet-hosting stars has been an
active field of study. Soon after the discovery of the first candidates,
claims were made that stars with planets displayed on average higher metal
contents (Gonzalez \cite{G97}) than other solar neighbourhood stars. A
number of subsequent independent studies with increasingly large stellar
samples have mostly confirmed the initial claims (e.g. Santos et al.
\cite{SIM03}). An important point to be made is that the determination of
chemical abundances, mostly carried out through detailed analysis of
spectroscopic data, is quite challenging (see Gonzalez \cite{G03} for a
complete review). As it has been shown for late-type stars (i.e. FGK
planet hosts), a strong degeneracy affecting the determination of metal
abundances is the correlation with effective temperature. Straightforward
estimations show that a systematic error of +100 K in $T_{\rm eff}$ (i.e.
1.5--2\% at the temperatures of FGK stars) results in metal abundances
being systematically overestimated by +0.06 dex ($\sim$15\%). Most studies
of stellar atmospheric parameters carry out multiple fits to derive
chemical compositions and effective temperatures from the spectra.
Although the aforementioned correlation would not alter the conclusions of
relative studies including both planet-hosting and non-planet hosting
stars, the metal richness of the stars would be systematically biased when
compared to the Sun. Another point worth making is the use by present
spectroscopic studies of solar line oscillator strengths for all late-type
stars (e.g. Gonzalez \& Laws \cite{GL00}). This might introduce systematic
errors for temperatures below and above that of the Sun that thus far have
not been addressed in detail.

The potential problems with spectroscopic analyses discussed above make a
completely independent temperature determination, for example using
photometry, very valuable. The absence of systematic effects when
comparing photometric and spectroscopic temperatures would strengthen the
case for the metal richness of planet-bearing stars and support the use of
solar oscillator strengths over the relevant spectral type range. However,
the determination of photometric temperatures for cool stars (below
7000~K) is not straightforward because most photometric systems are not
designed for such low temperatures. For example, although some efforts
have been made to extend the temperature range covered by Str\"omgren
calibrations down to late-type stars (Olsen \cite{O84}), most of the work
is still in a preliminary stage. Here we present a new approach, namely
the determination of effective temperatures from infrared (IR) photometry.
The underlying idea is similar to the Infra-Red Flux Method (IRFM),
proposed and implemented by Blackwell \& Shallis (\cite{BS77}), Blackwell
et al.  (\cite{BPS80}) and later Alonso et al. (\cite{AAM96}).

In this paper we briefly discuss the proposed approach and compare our
results with spectroscopic temperature determinations. In addition, the
analysis also yields an robust and accurate determination of the stellar
radius, provided the distance is known. With the release of the 2MASS All
Sky Catalog\footnote{http://www.ipac.caltech.edu/2mass}, which contains IR
photometry covering the entire sky, the proposed method has a wide
applicability, thus permitting accurate (a few percent) and effortless
determinations of temperatures and radii (important for transits) of
planet-hosting stars.

\section{The stellar sample}

Planet searching projects are currently so efficient that the number of
planet-hosting stars increases on a weekly basis. Our starting sample was
complete as of July 2003 and comprised 94 stars with planets, which makes
up a statistically significant number for our comparisons. The original
list was cross-matched with the 2MASS All Sky Catalog with a 100\% success
rate. In a next step, we excluded those stars with poor-quality photometry
in all of the three 2MASS bands, generally because of strong
saturation.\footnote{The 2MASS team managed to extract good quality
photometry from the wings of mildly saturated stars, which led to a
significant increase in dynamic range (see Cutri et al. \cite{CSV03}). In
general, good photometry was obtained for stars with $K>4$ mag.} As a
criterion, we rejected stars with errors in the $K$ band (the band with
best quality for bright stars) greater than 0.05 mag. In addition, two
stars were excluded for different reasons: HD 113020, classified as
spectral type M4, because its temperature is expected to be below 4000 K
and beyond the studied temperature range; and OGLE-TR56, the farthest
known planet-hosting star at 1.5 kpc, because of concerns with the
extinction correction. The final working sample is therefore composed of
81 planet-bearing stars, which are listed in Table \ref{tabTR}.

The 2MASS photometry was complemented with $V$ standard measurements
(together with their uncertainties) obtained from GCPD (Mermilliod et al.
\cite{MMH97}). Also from GCPD we compiled Str\"omgren photometry and
carried out an estimation of the reddening correction. All stars were
found to fall right on the standard relationships as defined by Crawford
(\cite{C75}) and Olsen (\cite{O84}, \cite{O88}), i.e. zero reddening as
expected from their nearby distances within 100 pc.

\section{Determination of temperatures and radii}

The use of IR photometry for determining effective temperatures was
initially proposed by Blackwell \& Shallis (\cite{BS77}). Their so-called
IRFM uses the ratio between the bolometric flux of the star and the
monochromatic at a given wavelength in the IR, both measured at the Earth,
as the observable quantity. This ratio is compared with a theoretical
estimate derived from atmosphere models, allowing the determination of the
effective temperature. The method we have developed follows a somewhat
different approach, namely, a fit of the stellar spectral energy
distribution from the optical ($V$) to the IR ($JHK$) with synthetic
photometry computed from atmosphere models. This wavelength range ensures
enough sensitivity to temperature, while being essentially immune to
intrinsic variations (mostly due to chromospheric activity) that would
strongly affect bluer passbands. Also importantly, the spectral energy
distribution in the optical/IR for cool stars is mostly insensitive to the
(often unknown) values of $\log g$ and $[Fe/H]$. In contrast to the IRFM,
our method does not require the use of a bolometric flux calibration, thus
reducing the risk of uncontrolled systematic errors, and permits the
determination of individual uncertainties. The algorithm will be discussed
in detail by Masana (\cite{M04}) but we briefly outline the procedure
here.

A critical pre-requisite to compare stellar and model fluxes is a
well-characterized photometric system and an accurate flux calibration.
Fortunately, the recent work by Cohen et al. (\cite{CMH03}, \cite{CWM03})
provides a set of consistent absolute flux calibrations in both the
optical (Landolt system) and IR (2MASS). To compute synthetic magnitudes
we employ the dense grid of Kurucz atmosphere
models\footnote{http://kurucz.harvard.edu/grids.html}, although tests with
other models (discussed below) reveal negligible differences. The
synthetic $VJHK$ photometry was calculated for each of the grid points as
a function of $T_{\rm eff}$, $[Fe/H]$, and $\log g$. We restricted our
calculations to the temperature interval between 4000 and 8000~K. The
upper limit is defined by the increasing dependence of the results on
$\log g$ and the lower limit is set by the decreasing performance of the
models because of molecular bands.

The algorithm itself is based on the minimisation of a $\chi^2$ function
defined from the difference between the observed and synthetic $VJHK$
magnitudes. The two adjustable parameters are $T_{\rm eff}$ and a
magnitude zeropoint (ZP), while $\log g$ and $[Fe/H]$ are fixed input data
(recall that interstellar absorption for our stars is negligible). The ZP
is the difference between the synthetic (star's surface)
and the observed magnitude (Earth surface), and it is directly
related to the semi-angular diameter ($\theta=10^{-0.2\,{\rm ZP}}$). The
input values of $[Fe/H]$ were taken from the spectroscopic determinations
and $\log g$ was roughly estimated from stellar models (in the $T_{\rm
eff}$--$L$ plane) after one iteration. Convergence towards the minimum
value of the reduced $\chi^2$ was very fast and reached values close to
unity.

Perhaps the greatest advantage of our method is that it yields the
individual uncertainties of both $T_{\rm eff}$ and $\theta$, calculated
from the covariance matrix, provided the involved errors are realistically
estimated. For the present sample, the error sources considered are: 1)
Observed magnitude errors, with an minimum 0.015 mag error in $V$ (usually
computed from the average of a few measurement and thus prone to error
underestimation); 2) Zero point uncertainties in the flux calibration
(1.5-1.7\%) from Cohen et al. (\cite{CMH03}, \cite{CWM03}); 3) Conservative
error bars of 0.1 dex in $[Fe/H]$ and 0.2 dex in $\log g$ dex\footnote{Our
method is essentially insensitive to the adopted value of $[Fe/H]$ and
$\log g$. Thus, uncertainties up to 0.3 dex in $[Fe/H]$ and 0.5 dex in
$\log g$ introduce uncertainties in $T_{\rm eff}$ below 0.5\%.}; 4) No
error was attributed to the model fluxes. We ran a number of comparison
with other (less dense) atmosphere model grids such as those by Castelli
et al. (\cite{CGK97}) and the NextGen models by Hauschildt et al.
\cite{HAB99}). Our tests were extremely satisfactory and yielded average
temperature differences below 15 K ($\sim$0.3\%) in all cases. Thus,
systematic errors introduced by atmosphere models are likely to be
negligible.

The procedure described above yields two basic parameters: The
best-fitting effective temperature and semi-angular diameter (i.e.
radius/distance). With a known distance, the latter can be transformed
into a true radius measurement. The results for our stellar sample are
listed in Table \ref{tabTR}. The relative accuracy of the results is
0.5--1.3\% in effective temperature and 0.9--2.4\% in semi-angular
diameter. Since the planet-hosting stars in the sample are generally
nearby, their Hipparcos distances are very accurate. This results in
relative uncertainties in the individual radii in the range of 1.3--4.7\%
for 90\% of the stars.

\begin{table*}[!t]
\centering
\caption[]{Effective temperatures, semi-angular diameters ($\theta\equiv$
radius/distance) and radii, with their uncertainties, of planet-hosting
stars as resulting from the algorithm discussed in this paper.}
\label{tabTR}
\scriptsize
\begin{tabular}{rcccrcccrccc}
\hline
\hline
\multicolumn{1}{c}{HD} &
$T_{\rm eff}$ &
$\theta$ &
$R$ &
\multicolumn{1}{c}{HD} &
$T_{\rm eff}$ &
$\theta$ &
$R$ &
\multicolumn{1}{c}{HD/} &
$T_{\rm eff}$ &
$\theta$ &
$R$ \\
 &
(K) & 
(mas) &
(R$_{\odot}$) &  
 &
(K) & 
(mas) &
(R$_{\odot}$) &  
\multicolumn{1}{c}{BD} &
(K) & 
(mas) & 
(R$_{\odot}$) \\
\hline
   142 & $\!\!\!$6304$\pm$65  & $\!\!\!$0.255$\pm$0.005  & $\!\!\!$1.404$\pm$0.029  & 68988 & $\!\!\!$5911$\pm$47  & $\!\!\!$0.094$\pm$0.001  & $\!\!\!$1.183$\pm$0.044  &145675                   & $\!\!\!$5357$\pm$32  & $\!\!\!$0.251$\pm$0.003  & $\!\!\!$0.978$\pm$0.013  \\
  1237 & $\!\!\!$5600\hfill45 & $\!\!\!$0.224\hfill0.003 & $\!\!\!$0.849\hfill0.013 & 72659 & $\!\!\!$5912\hfill52 & $\!\!\!$0.132\hfill0.002 & $\!\!\!$1.454\hfill0.057 &147513                   & $\!\!\!$5978\hfill70 & $\!\!\!$0.337\hfill0.008 & $\!\!\!$0.933\hfill0.024 \\
  2039 & $\!\!\!$5940\hfill50 & $\!\!\!$0.063\hfill0.001 & $\!\!\!$1.227\hfill0.085 & 73256 & $\!\!\!$5480\hfill39 & $\!\!\!$0.121\hfill0.002 & $\!\!\!$0.950\hfill0.017 &150706                   & $\!\!\!$5961\hfill47 & $\!\!\!$0.160\hfill0.002 & $\!\!\!$0.938\hfill0.016 \\
  3651 & $\!\!\!$5394\hfill56 & $\!\!\!$0.347\hfill0.008 & $\!\!\!$0.829\hfill0.020 & 73526 & $\!\!\!$5661\hfill46 & $\!\!\!$0.073\hfill0.001 & $\!\!\!$1.491\hfill0.106 &162020                   & $\!\!\!$4745\hfill26 & $\!\!\!$0.116\hfill0.001 & $\!\!\!$0.783\hfill0.037 \\
  4208 & $\!\!\!$5688\hfill46 & $\!\!\!$0.125\hfill0.002 & $\!\!\!$0.879\hfill0.025 & 74156 & $\!\!\!$6006\hfill50 & $\!\!\!$0.118\hfill0.002 & $\!\!\!$1.638\hfill0.092 &168443                   & $\!\!\!$5584\hfill40 & $\!\!\!$0.194\hfill0.002 & $\!\!\!$1.582\hfill0.051 \\
  4203 & $\!\!\!$5611\hfill43 & $\!\!\!$0.084\hfill0.001 & $\!\!\!$1.411\hfill0.112 & 75289 & $\!\!\!$6151\hfill52 & $\!\!\!$0.200\hfill0.003 & $\!\!\!$1.243\hfill0.022 &168746                   & $\!\!\!$5577\hfill44 & $\!\!\!$0.121\hfill0.002 & $\!\!\!$1.124\hfill0.045 \\
  6434 & $\!\!\!$5829\hfill51 & $\!\!\!$0.122\hfill0.002 & $\!\!\!$1.060\hfill0.030 & 75732 & $\!\!\!$5338\hfill53 & $\!\!\!$0.343\hfill0.008 & $\!\!\!$0.925\hfill0.023 &169830                   & $\!\!\!$6325\hfill54 & $\!\!\!$0.230\hfill0.003 & $\!\!\!$1.799\hfill0.060 \\
  8574 & $\!\!\!$6056\hfill51 & $\!\!\!$0.145\hfill0.002 & $\!\!\!$1.380\hfill0.049 & 76700 & $\!\!\!$5670\hfill46 & $\!\!\!$0.106\hfill0.001 & $\!\!\!$1.367\hfill0.053 &177830                   & $\!\!\!$4841\hfill31 & $\!\!\!$0.257\hfill0.003 & $\!\!\!$3.268\hfill0.091 \\
 10697 & $\!\!\!$5620\hfill42 & $\!\!\!$0.256\hfill0.004 & $\!\!\!$1.796\hfill0.051 & 80606 & $\!\!\!$5605\hfill43 & $\!\!\!$0.074\hfill0.001 & $\!\!\!$0.927\hfill0.304 &178911                   & $\!\!\!$5630\hfill42 & $\!\!\!$0.114\hfill0.001 & $\!\!\!$1.145\hfill0.502 \\
 12661 & $\!\!\!$5732\hfill43 & $\!\!\!$0.144\hfill0.002 & $\!\!\!$1.154\hfill0.035 & 82943 & $\!\!\!$5992\hfill48 & $\!\!\!$0.193\hfill0.002 & $\!\!\!$1.138\hfill0.028 &179949                   & $\!\!\!$6202\hfill52 & $\!\!\!$0.205\hfill0.003 & $\!\!\!$1.193\hfill0.030 \\
 13445 & $\!\!\!$5253\hfill54 & $\!\!\!$0.333\hfill0.008 & $\!\!\!$0.781\hfill0.019 & 83443 & $\!\!\!$5508\hfill41 & $\!\!\!$0.110\hfill0.001 & $\!\!\!$1.034\hfill0.034 &186427                   & $\!\!\!$5729\hfill42 & $\!\!\!$0.251\hfill0.003 & $\!\!\!$1.154\hfill0.019 \\
 16141 & $\!\!\!$5785\hfill47 & $\!\!\!$0.185\hfill0.003 & $\!\!\!$1.428\hfill0.047 & 89744 & $\!\!\!$6189\hfill55 & $\!\!\!$0.260\hfill0.004 & $\!\!\!$2.181\hfill0.055 &187123                   & $\!\!\!$5849\hfill45 & $\!\!\!$0.114\hfill0.001 & $\!\!\!$1.179\hfill0.032 \\
 17051 & $\!\!\!$6286\hfill80 & $\!\!\!$0.296\hfill0.007 & $\!\!\!$1.097\hfill0.029 & 92788 & $\!\!\!$5705\hfill44 & $\!\!\!$0.152\hfill0.002 & $\!\!\!$1.055\hfill0.033 &190228                   & $\!\!\!$5282\hfill37 & $\!\!\!$0.191\hfill0.002 & $\!\!\!$2.555\hfill0.065 \\
 20367 & $\!\!\!$6055\hfill52 & $\!\!\!$0.202\hfill0.003 & $\!\!\!$1.182\hfill0.029 &106252 & $\!\!\!$5889\hfill51 & $\!\!\!$0.136\hfill0.002 & $\!\!\!$1.092\hfill0.041 &190360                   & $\!\!\!$5639\hfill53 & $\!\!\!$0.326\hfill0.006 & $\!\!\!$1.113\hfill0.023 \\
 23079 & $\!\!\!$6030\hfill52 & $\!\!\!$0.149\hfill0.002 & $\!\!\!$1.106\hfill0.022 &108147 & $\!\!\!$6245\hfill50 & $\!\!\!$0.144\hfill0.002 & $\!\!\!$1.193\hfill0.028 &192263                   & $\!\!\!$5013\hfill33 & $\!\!\!$0.178\hfill0.002 & $\!\!\!$0.760\hfill0.019 \\
 23596 & $\!\!\!$6071\hfill51 & $\!\!\!$0.137\hfill0.002 & $\!\!\!$1.530\hfill0.057 &108874 & $\!\!\!$5568\hfill43 & $\!\!\!$0.084\hfill0.001 & $\!\!\!$1.232\hfill0.082 &195019                   & $\!\!\!$5666\hfill39 & $\!\!\!$0.191\hfill0.002 & $\!\!\!$1.538\hfill0.041 \\
 28185 & $\!\!\!$5704\hfill48 & $\!\!\!$0.122\hfill0.002 & $\!\!\!$1.041\hfill0.038 &114386 & $\!\!\!$4883\hfill29 & $\!\!\!$0.124\hfill0.002 & $\!\!\!$0.750\hfill0.028 &196050                   & $\!\!\!$5894\hfill49 & $\!\!\!$0.129\hfill0.002 & $\!\!\!$1.306\hfill0.037 \\
 30177 & $\!\!\!$5633\hfill47 & $\!\!\!$0.096\hfill0.002 & $\!\!\!$1.130\hfill0.049 &114762 & $\!\!\!$5899\hfill50 & $\!\!\!$0.146\hfill0.002 & $\!\!\!$1.273\hfill0.053 &209458                   & $\!\!\!$6088\hfill56 & $\!\!\!$0.113\hfill0.002 & $\!\!\!$1.145\hfill0.049 \\
 33636 & $\!\!\!$5967\hfill46 & $\!\!\!$0.159\hfill0.002 & $\!\!\!$0.984\hfill0.032 &114783 & $\!\!\!$5157\hfill33 & $\!\!\!$0.179\hfill0.002 & $\!\!\!$0.788\hfill0.018 &210277                   & $\!\!\!$5578\hfill43 & $\!\!\!$0.232\hfill0.003 & $\!\!\!$1.061\hfill0.021 \\
 37124 & $\!\!\!$5577\hfill43 & $\!\!\!$0.140\hfill0.002 & $\!\!\!$1.004\hfill0.039 &114729 & $\!\!\!$5853\hfill49 & $\!\!\!$0.194\hfill0.003 & $\!\!\!$1.462\hfill0.050 &213240                   & $\!\!\!$5939\hfill46 & $\!\!\!$0.175\hfill0.002 & $\!\!\!$1.533\hfill0.039 \\
 39091 & $\!\!\!$6006\hfill64 & $\!\!\!$0.290\hfill0.006 & $\!\!\!$1.138\hfill0.025 &121504 & $\!\!\!$5994\hfill47 & $\!\!\!$0.122\hfill0.002 & $\!\!\!$1.167\hfill0.035 &216435                   & $\!\!\!$5992\hfill50 & $\!\!\!$0.245\hfill0.004 & $\!\!\!$1.753\hfill0.042 \\
 38529 & $\!\!\!$5578\hfill57 & $\!\!\!$0.305\hfill0.007 & $\!\!\!$2.784\hfill0.108 &128311 & $\!\!\!$4936\hfill28 & $\!\!\!$0.214\hfill0.002 & $\!\!\!$0.762\hfill0.013 &216437                   & $\!\!\!$5942\hfill56 & $\!\!\!$0.247\hfill0.004 & $\!\!\!$1.409\hfill0.030 \\
 40979 & $\!\!\!$6223\hfill55 & $\!\!\!$0.165\hfill0.002 & $\!\!\!$1.183\hfill0.032 &130322 & $\!\!\!$5464\hfill39 & $\!\!\!$0.123\hfill0.002 & $\!\!\!$0.787\hfill0.030 &217014                   & $\!\!\!$5827\hfill52 & $\!\!\!$0.343\hfill0.006 & $\!\!\!$1.133\hfill0.022 \\
 46375 & $\!\!\!$5324\hfill37 & $\!\!\!$0.140\hfill0.002 & $\!\!\!$1.005\hfill0.036 &134987 & $\!\!\!$5844\hfill42 & $\!\!\!$0.214\hfill0.002 & $\!\!\!$1.178\hfill0.028 &217107                   & $\!\!\!$5694\hfill44 & $\!\!\!$0.260\hfill0.004 & $\!\!\!$1.103\hfill0.022 \\
 49674 & $\!\!\!$5642\hfill41 & $\!\!\!$0.111\hfill0.001 & $\!\!\!$0.968\hfill0.040 &136118 & $\!\!\!$6136\hfill57 & $\!\!\!$0.154\hfill0.002 & $\!\!\!$1.736\hfill0.071 &222582                   & $\!\!\!$5831\hfill50 & $\!\!\!$0.123\hfill0.002 & $\!\!\!$1.107\hfill0.044 \\
 50554 & $\!\!\!$6020\hfill49 & $\!\!\!$0.167\hfill0.002 & $\!\!\!$1.117\hfill0.035 &141937 & $\!\!\!$5881\hfill47 & $\!\!\!$0.146\hfill0.002 & $\!\!\!$1.052\hfill0.039 &\llap{$-$02} 5917\rlap{B}& $\!\!\!$5542\hfill51 & $\!\!\!$0.094\hfill0.002 & $\!\!\!$1.100\hfill0.120 \\
 52265 & $\!\!\!$6107\hfill50 & $\!\!\!$0.208\hfill0.003 & $\!\!\!$1.258\hfill0.031 &143761 & $\!\!\!$5825\hfill57 & $\!\!\!$0.354\hfill0.006 & $\!\!\!$1.327\hfill0.027 &\llap{$-$10} 3166        & $\!\!\!$5315\hfill40 & $\!\!\!$0.052\hfill0.001 & $\!\!\!$       --        \\
\hline
\end{tabular}
\end{table*}

\section{Discussion}

\begin{figure}[!t]
\centering
\includegraphics[width=7.0cm]{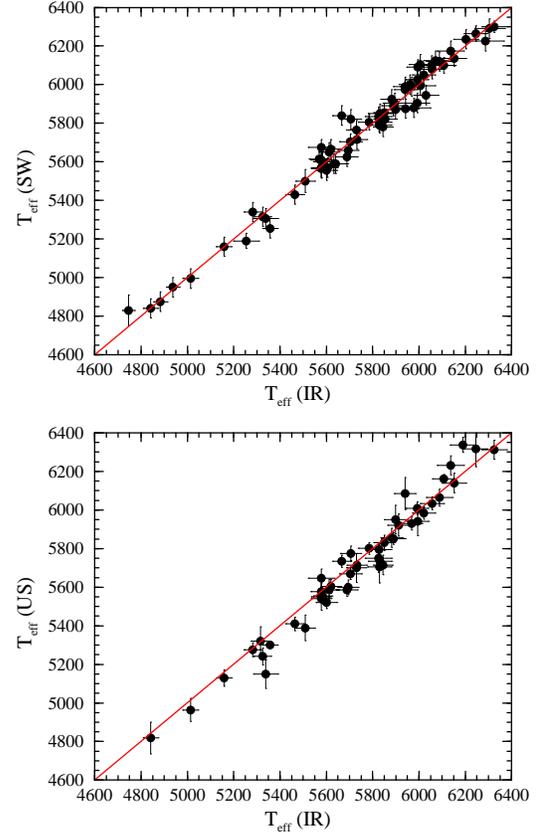}
\caption{Comparison of effective temperatures determined using our method
(IR) and temperatures determined spectroscopically by the Swiss (SW) and
US groups (see text).}
\label{figCTef}
\end{figure}

With the temperatures from IR photometry in hand, we carried out a
comparison with spectroscopic determinations to assess their mutual
agreement. We considered two comprehensive and independent spectroscopic
analyses, following, however, similar approaches:  The Swiss group (Santos
et al. \cite{SIM01}, \cite{SIM03}) and the US group (Gonzalez \cite{G97},
\cite{G98}, \cite{G99}; Gonzalez \& Laws \cite{GL00}; Gonzalez et al.
\cite{GLT01}; Feltzing \& Gonzalez \cite{FG01}; Laws et al. \cite{LGW03}).
A cross-match of the IR and spectroscopic samples yields a total of 69
stars in common with the Swiss group and 49 with the US group. The
resulting comparisons are illustrated in Fig. \ref{figCTef}, where the
observational data are plotted with their error bars. As can be seen, the
agreement among the three temperature determinations is very remarkable.
The mean difference between the IR-based and the Swiss group temperatures
is $<{T_{\rm eff}}_{\rm IR}-{T_{\rm eff}}_{\rm SW}>=-4.2\pm51.4$~K, with
no hint of any systematic trends neither as a function of temperature,
metallicity or surface gravity. The difference with the US group yields a
value of $<{T_{\rm eff}}_{\rm IR}-{T_{\rm eff}}_{\rm US}>=+20.5\pm65.9$~K,
with a low significance trend of larger IR temperatures at the low $T_{\rm
eff}$ end and smaller IR temperatures at the high $T_{\rm eff}$ end. No
systematic trends are seen as a function of metallicity or surface
gravity. Interestingly, in both cases the scatter of the differences is
entirely consistent with the quoted error bars, which indicates that our
procedure yields realistic uncertainties.

A valuable independent test of our analysis can be carried out by
comparing the semi-angular diameters with observational
determinations. Unfortunately, this is not possible at this point because
all the stars with available empirical angular diameter measurements are
very bright (see Alonso et al. \cite{AAM94} for a compilation) and do not
have accurate 2MASS photometry. Luckily, there is one star in our sample
with an accurate empirical radius measurement and this is HD 209458, which
undergoes planetary transits. Generally it is very difficult to
disentangle the stellar and the planetary radii effects when analyzing
transit light curves. However, Brown et al. (\cite{BCG01}) were able to
determine both the stellar radius and the planetary radius simultaneously
from a high precision transit light curve of HD 209458 obtained with the
Hubble Space Telescope. Their empirically-determined radius is
$1.146\pm0.050$~R$_{\odot}$, which can be compared with our estimate of
$1.145\pm0.049$~R$_{\odot}$ from the fit to the star's optical/IR energy
distribution. The agreement is excellent (owing the almost null difference
to small number statistics) and the error bars of both measurements are
alike.

These important results indicate that our temperature and radius
determination method is robust and with a very similar accuracy to the one
possibly achievable with the best available spectroscopy and photometry
today.

For completeness, the effective temperatures in Table \ref{tabTR} were
also compared with those computed from photometric calibrations. A
detailed discussion is left for a forthcoming publication, but we shall
briefly review here some of the most relevant results. Our tests focused
on the Str\"omgren calibrations of Olsen (\cite{O84}) and the IRFM as
implemented by Alonso et al. (\cite{AAM96}). The comparisons show that the
$(b-y)$ calibration of Olsen (\cite{O84}) yields effective temperatures
that are systematically lower than those resulting from our method, with
the mean difference being 168$\pm$103~K. Interestingly, a marked trend was
found when plotting the differences as a function of metallicity. It seems
that the origin of the discrepancies can be ascribed to the stellar sample
used to derive the Str\"omgren calibrations, since only one third of
Olsen's reference stars have super-solar metallicities. In contrast, the
comparison with the temperatures derived from the IRFM calibration of
Alonso et al. (\cite{AAM96}) reveals no systematic trends but they are on
average 76$\pm$42~K smaller than our determinations. Given the similarity
of both methods and input data, the discrepancy can tentatively be
attributed to the photometric transformations employed or to the
bolometric flux calibration used by Alonso et al. (\cite{AAM96}).

\section{Conclusions}

The conclusions of our study are twofold. First, we have compared the
effective temperature determinations for planet-bearing stars from two
completely independent approaches with similar accuracies, namely detailed
spectroscopic analysis and IR photometry. The results indicate an
excellent agreement in the entire temperature range, which confidently
rules out the possibility of systematic effects in spectroscopic
metallicity determinations and supports the use of solar line oscillator
strengths. Second, the method presented, consisting in a fit to the
observed $VJHK$ magnitudes using synthetic magnitudes, has proved its
reliability, yielding accurate ($\sim$1\%) and cost-effective
temperatures. As a bonus, the analysis also provides determinations of the
semi-angular diameters and, eventually, the stellar radii. The resulting
radius accuracy of a few percent (for nearby stars) could be extremely
useful to break the strong degeneracy between the radii of the planet and
the star when analysing transit light curves.

\begin{acknowledgements} 
I. R. and E. M. acknowledge support from the Spanish MCyT through grant
AyA2000-0937. This publication makes use of data products from the Two
Micron All Sky Survey, which is a joint project of the University of
Massachusetts and the Infrared Processing and Analysis Center/California
Institute of Technology, funded by the National Aeronautics and Space
Administration and the National Science Foundation.
\end{acknowledgements}

\end{document}